\documentclass[aps,nofootinbib,superscriptaddress,twocolumn]{revtex4}
\usepackage{graphicx}

\newcommand{\bnmr}{$\beta$-NMR}
\newcommand{\Li}{$^8$Li}
\newcommand{\Mn}{Mn$_{12}$}

\begin{document}
\title{Local Magnetic Properties of a Monolayer of \Mn\ Single Molecule
 Magnets}

\author{Z.~Salman}\
\email{z.salman1@physics.ox.ac.uk}
\affiliation{Clarendon Laboratory, Department of
  Physics, Oxford University, Parks Road, Oxford OX1 3PU, UK}
\affiliation{ISIS Facility, Rutherford Appleton Laboratory, Chilton,
  Oxfordshire, OX11 0QX, UK}
\affiliation{TRIUMF, 4004 Wesbrook Mall, Vancouver, BC, Canada, V6T
  2A3}
\author{K.~H.~Chow}
\affiliation{Department of Physics, University of Alberta, Edmonton,
  AB, Canada, T6G 2G7}
\author{R.~I.~Miller}
\affiliation{TRIUMF, 4004 Wesbrook Mall, Vancouver, BC, Canada, V6T 2A3}
\author{A.~Morello}
\affiliation{Department of Physics and Astronomy, University of
  British Columbia, Vancouver, BC, Canada V6T 1Z1}
\author{T.~J.~Parolin}
\affiliation{Department of Chemistry, University of British Columbia,
  Vancouver, BC, Canada V6T 1Z1}
\author{M.~D.~Hossain}
\author{T.~A.~Keeler}
\affiliation{Department of Physics and Astronomy, University of
  British Columbia, Vancouver, BC, Canada V6T 1Z1}
\author{C.~D.~P.~Levy}
\affiliation{TRIUMF, 4004 Wesbrook Mall, Vancouver, BC, Canada, V6T 2A3}
\author{W.~A.~MacFarlane}
\affiliation{Department of Chemistry, University of British Columbia,
  Vancouver, BC, Canada V6T 1Z1}
\author{G.~D.~Morris}
\affiliation{TRIUMF, 4004 Wesbrook Mall, Vancouver, BC, Canada, V6T 2A3}
\author{H.~Saadaoui}
\author{D.~Wang}
\affiliation{Department of Physics and Astronomy, University of
  British Columbia, Vancouver, BC, Canada V6T 1Z1}
\author{R.~Sessoli}
\affiliation{Dipartimento di Chimica, Universit\'a di Firenze \&
  INSTM, via della Lastruccia 3, 50019 Sesto Fiorentino, Italy}
\author{G.~G.~Condorelli}
\affiliation{Dipartimento di Scienze Chimiche, Universit\'a di Catania
\& INSTM UdR di Catania, viale A. Doria 6, 95125 Catania, Italy}
\author{R.~F.~Kiefl}
\affiliation{TRIUMF, 4004 Wesbrook Mall, Vancouver, BC, Canada, V6T 2A3}
\affiliation{Department of Physics and Astronomy, University of
  British Columbia, Vancouver, BC, Canada V6T 1Z1}

\begin{abstract}
  The magnetic properties of a monolayer of \Mn\ single molecule
  magnets grafted onto a Si substrate have been investigated using
  depth-controlled $\beta$-detected nuclear magnetic resonance.  A low
  energy beam of spin polarized radioactive \Li\ was used to probe the
  local static magnetic field distribution near the \Mn\ monolayer in
  the Si substrate. The resonance linewidth varies strongly as a
  function of implantation depth as a result of the magnetic dipolar
  fields generated by the \Mn\ electronic magnetic moments. The
  temperature dependence of the linewidth indicates that the magnetic
  properties of the \Mn\ moments in this low dimensional configuration
  differ from bulk \Mn.
\end{abstract}
\maketitle

Single molecule magnets (SMMs)\cite{Sessoli03ACIE} are molecules which
contain a small number of magnetic ions with large magnetic
interactions between them ($J \sim 10-100$~K). The magnetic cores of
each molecule are surrounded by organic or inorganic ligands. Since
the molecules are magnetically isolated, they form at low temperature
a lattice of very weakly interacting spins. Practical application of
SMMs as molecular scale units for information storage
\cite{Stamp96LTP,Joachim00N} or ``qubits'' for quantum computation
\cite{Divincenzo95QTM,Tejada01N,Troiani05PRL,Troiani05PRL2} requires
addressing individual molecules, which may be realized in principle by
depositing a monolayer of molecules on a suitable substrate.  Methods
to deposit suitably derivatized \Mn-type clusters on gold
\cite{Cornia03ACIE,Mannini05NL,Zobbi05CC,Naitabdi05AM} and Si
\cite{Condorelli04ACIE,Fleury05CC} have been developed recently,
opening up exciting possibilities for applications of SMMs for
information storage on a single molecule and for the investigation of
the quantum behavior of isolated spins, such as quantum tunnelling of
the magnetization (QTM)
\cite{Friedman96PRL,Thomas96N,Sangregorio97PRL,Sessoli03ACIE},
topological quantum phase interference
\cite{Wernsdorfer99S,Wernsdorfer00EL}, and quantum coherence
\cite{Hill03S,delBarco04PRL,Morello06PRL}. Unfortunately, the small
quantity of magnetic material in the case of a monolayer (or
sub-monolayer\cite{Naitabdi05AM}) implies that it is virtually
impossible to accurately determine magnetic properties with
conventional bulk techniques, such as SQUID magnetometry or
conventional nuclear magnetic resonance (NMR). However, a new
technique, namely depth-resolved $\beta$-detected NMR (\bnmr), which
has $\approx 10^{13}$ orders of magnitude higher sensitivity compared
to conventional NMR, is well-suited for studying such
systems\cite{Keeler06PB,Morris04PRL,Salman06PRL,Salman07PRB,Parolin07PRL}.

In this paper we report \bnmr\ measurements of the magnetic moment of
\Mn\ molecules which are grafted as a monolayer on a Si substrate.
The experiments were performed using a low energy beam of highly
polarized radioactive \Li, implanted into the Si substrate just below
the \Mn\ monolayer.  The strength and distribution of the magnetic
dipolar fields from the \Mn\ moments determines the shape of the \Li\
NMR resonance. Interestingly, the temperature dependence of the signal
deviates significantly from the measured magnetization for bulk
\Mn. This is evidence that the interactions characterizing \Mn\ in
this 2D configuration are different from the bulk.

\begin{figure*}[htb]
  \centerline{\includegraphics[height=7.0cm]{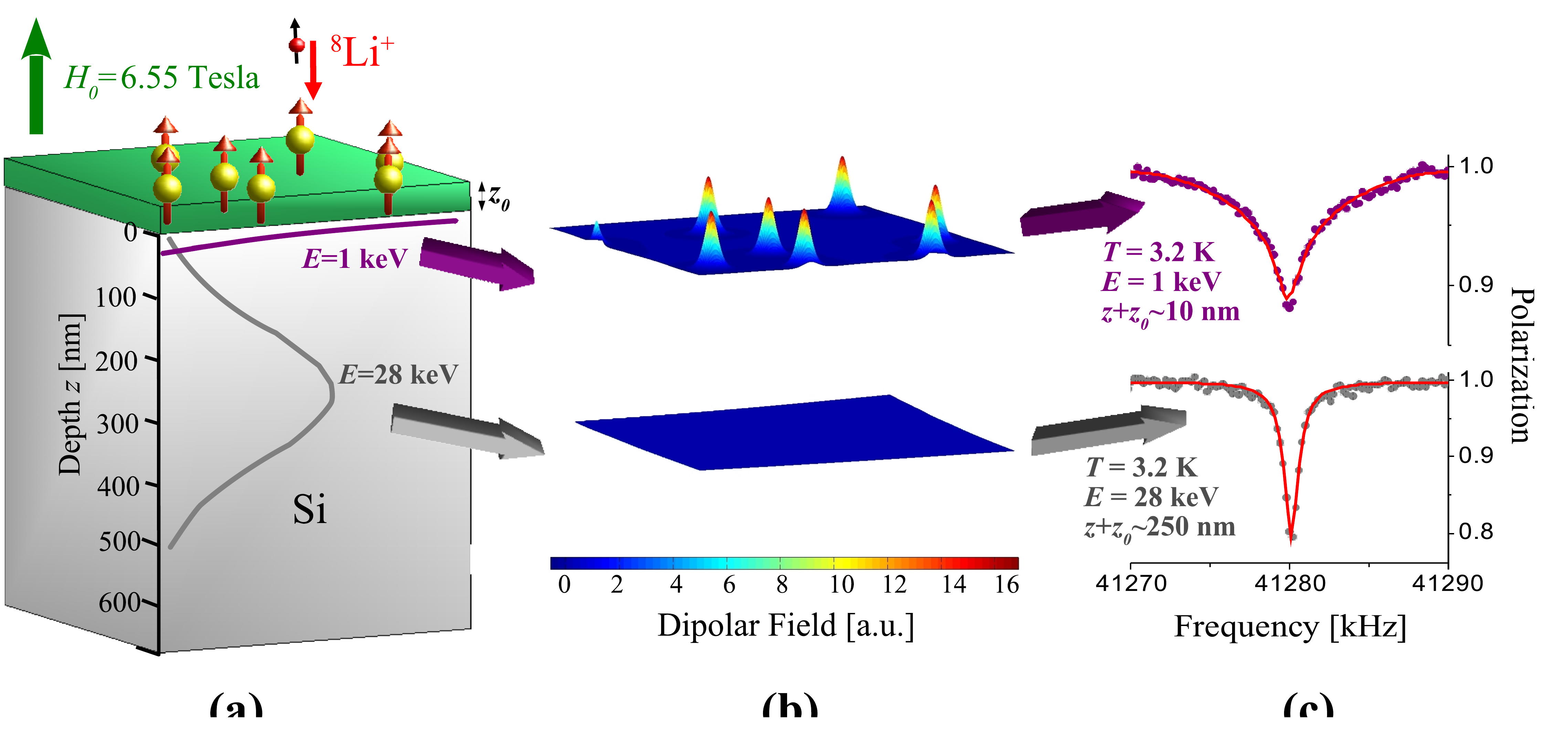}}
  \caption{(a) A schematic of sample {\bf 1} where the \Mn\ molecules
  are grafted on a Si substrate. The stopping profiles of \Li\ in Si
  at $E=1$ and $28$~keV are also shown (purple and grey lines
  respectively). (b) The simulated dipolar fields from the \Mn\
  monolayer calculated near the surface (top) and deep within (bottom)
  the Si substrate (arbitrary units). (c) The measured \bnmr\ spectra
  from sample {\bf 1} in an applied magnetic field $H_0=6.55$ Tesla at
  $T=3.2$~K. The top spectrum is for $E=1$ keV and the bottom for
  $E=28$ keV. The solid lines are fits to the calculated resonance
  line-shape (see text).}
  \label{Mn12-3D}
\end{figure*}
Magnetic resonance techniques have been used extensively to study the
magnetic properties of SMMs in the bulk. In particular, conventional
NMR\cite{Lascialfari98PRB,Lascialfari98PRL,Furukawa03PRB,Goto03PRB,Morello04PRL}
and muon spin
relaxation\cite{Lascialfari98PRB,Lascialfari98PRL,Salman02PRB,Salman00PB}
($\mu$SR) have been used to measure the molecular spin dynamics in
both the thermally activated regime as well as the quantum tunneling
regime. \bnmr\ is a closely related technique, where one measures the
nuclear magnetic resonance and relaxation of \Li, a spin $I=2$ nucleus
with a small electric quadrupole moment $Q=+31$~mB and gyromagnetic
ratio $\gamma=6.301$ MHz/T. The radioactive \Li$^+$ beam is produced
at the isotope separator and accelerator (ISAC) at TRIUMF. It is then
polarized using a collinear optical pumping method, and implanted into
the sample. Since the implantation energy can be varied between
$0.9-28$~keV, corresponding to an average implantation depth in Si of
$1-250$~nm, depth resolved \bnmr\ measurements are possible. As in any
form of NMR, the time evolution of the nuclear polarization is the
quantity of interest. It can be measured through the $\beta$-decay
asymmetry, where an electron is emitted preferentially opposite to the
direction of the nuclear polarization at the time of decay
\cite{Crane01PRL} and detected by appropriately positioned
scintillation counters.  As noted above, this method of detection is
dramatically more sensitive than conventional NMR and makes \bnmr\
suitable for studies of ultra-thin films and nano-structures
\cite{Morris04PRL,Keeler06PB}. The nuclear resonance in a static
magnetic field, $\mathbf{H_0}=H_0 \hat{z}$, can be detected by
measuring the time averaged nuclear polarization along $\hat{z}$,
$p_z(\nu)$, as a function of the frequency $\nu$ of a small ($\sim 1$
G) oscillating perpendicular magnetic field, $\mathbf{H_1}(t)=H_1
\cos(2 \pi \nu t) \hat{x}$. A loss of polarization occurs when $\nu$
matches the Larmor frequency of the nuclear spins of the \Li, a value
that is given by the product of $\gamma$ and the local field it
experiences. Hence, the position and shape of the resonance(s) signals
provide detailed information on the distribution of static local
magnetic fields.

The experiments reported here were performed on two different samples.
Sample {\bf 1} was prepared using a three-step
process\cite{Condorelli04ACIE}: 1) grafting of methyl ester of
10-undecanoic acid on a H-terminated Si(100) substrate, 2) hydrolysis
of the ester group, and 3) ligand exchange between
[Mn$_{12}$O$_{12}$(OAc)$_{16}$(H$_2$O)$_4] \cdot $H$_2$O$\cdot 2$AcOH
and the grafted undecanoic acid to anchor the \Mn\ SMMs to the organic
layer. A schematic of sample {\bf 1} is shown in
Fig.~\ref{Mn12-3D}(a). Sample {\bf 2} is an identically prepared Si
substrate, i.e. following step 1 only. It is used as a control sample
in order to confirm that the effects measured in {\bf 1} are solely
due to the \Mn. The samples were mounted in an ultra high vacuum (UHV)
environment on a cold finger cryostat. The resonance lines of \Li\
were measured at various temperatures and implantation energies in
both samples in an external magnetic field $H_0=6.55$ Tesla,
perpendicular to the Si surface.

The \bnmr\ spectra were measured by implanting the \Li\ beam at
different energies in the Si substrate {\em below} the \Mn\
monolayer. An example of the stopping profile of the implanted \Li\ at
two different energies is shown in Fig.~\ref{Mn12-3D}(a). At
$E=1$~keV, where most of the \Li\ stop within $10$~nm of the Si
surface, the dipolar field from the \Mn\ moments is large, as
illustrated in Fig.~\ref{Mn12-3D}(b). However, at $E=28$~keV the
average \Li\ implantation depth is $\sim 250$~nm, and the dipolar
field at this depth is negligible; hence, the local field experienced
by the \Li\ is simply the applied uniform $\mathbf{H_0}$. As a result
the measured resonance line at $1$~keV is significantly broadened
compared to that measured at $28$~keV, as clearly seen in
Fig.~\ref{Mn12-3D}(c) at $T=3.2$~K. Furthermore, the resonance
measured in sample {\bf 2} at $E=28$~keV and $T=3.2$~K is identical to
that measured in sample {\bf 1} under the same conditions, and the
broadening observed in sample {\bf 2} is much smaller at
$E=1$~keV. This demonstrates that low energy \bnmr\ spectroscopy is
sensitive to the magnetization of the \Mn\ monolayer.  In particular,
the \Li\ nuclei implanted into sample {\bf 1} at low $E$, and hence
stopping close to the \Mn\ molecules, experience a large distribution
of magnetic fields, which is attributed to the dipolar fields from the
\Mn\ monolayer.

\begin{figure}[h]
\centerline{\includegraphics[width=\columnwidth]{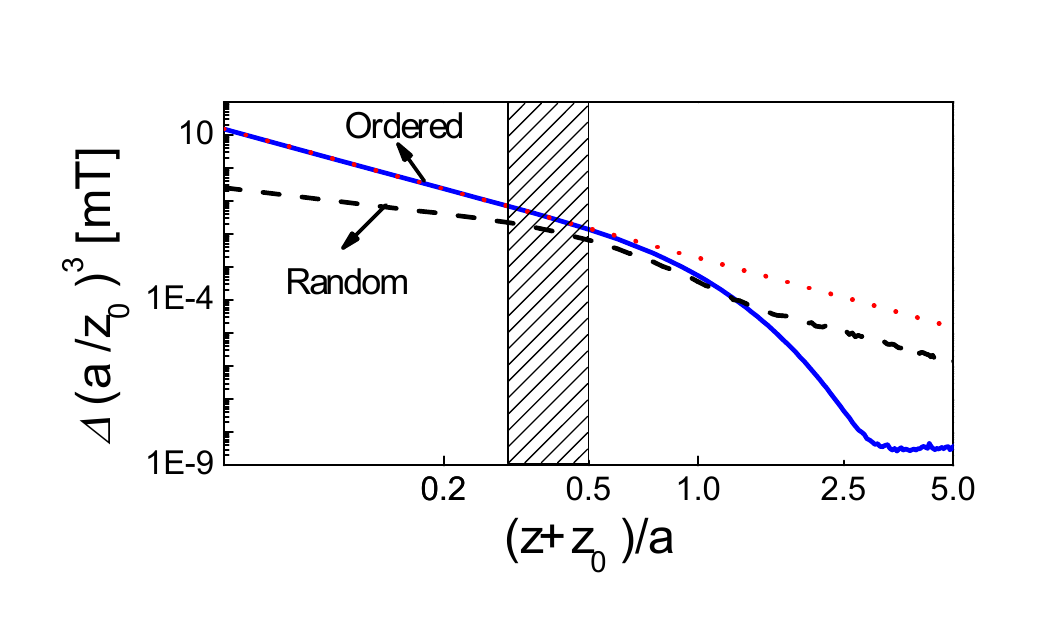}}
\caption{The simulated value of $\Delta$ as a function of distance
  from a monolayer of $1 \mu_B$ moments for square and random lattices
  (solid and dashed lines respectively). $\Delta$ near the monolayer
  follows Eq.~(\ref{Bmax}) with $\alpha=3.0$ (dotted line). The
  hatched area indicates the depth where $\alpha$ deviates from its
  asymptotic value.}
\label{Square}
\end{figure}

The observed resonance broadening, depicted in Fig.~\ref{Mn12-3D}(c),
can be described in terms of dipolar fields from the \Mn\ moments
$\langle \mathbf{m} \rangle =m\hat{z}$, which are preferentially
aligned by $\mathbf{H_0}$. For discussion purposes, let us start by
assuming that the \Mn\ moments are arranged in a 2 dimensional square
lattice with a lattice constant $a$ at $z=-z_0$, where the Si
substrate surface is assumed to be at $z=0$.  The $z$ component of the
total dipolar field, experienced by a \Li\ at $\mathbf{R}=x \hat{x}+y
\hat{y} + z \hat{z}$, due to moments $m_i$ at $\mathbf{R_i}=X_i
\hat{x}+ Y_i \hat{y} - z_0 \hat{z}$ is
\begin{equation}
  H_z^d(z)=\sum_i \frac{\mu_{0}m}{4\pi r_i^{3}}\left(\frac{3z^{2}}{r_i^{2}}-1\right),
\end{equation}
where $\mathbf{r_i}=\mathbf{R}-\mathbf{R_i}$. As we shall see below,
it is useful for calculation purposes to parameterize the width of the
dipolar field distribution experienced by a \Li\ stopping at a depth
$z$ with a reasonable analytical function. Simulations indicate that
the dipolar fields from the \Mn\ monolayer decay in the Si according
to a power law:
\begin{equation} \label{Bmax}
\Delta(z) = \Delta_0 \left(1+\frac{z}{z_0}\right)^{-\alpha}
\end{equation}
where $\Delta_0$ is the width of dipolar field distribution at the
surface of the Si substrate ($z=0$), $z_0$ is the distance between the
monolayer and the Si surface and $\alpha$ is a parameter describing
the decay of dipolar field in the substrate as a function of
depth. Note that $\Delta_0$ is proportional to the magnetic moment
$m$. In Fig. ~\ref{Square} we plot the results of the simulation for
$\Delta$ as a function of distance from the plane of the monolayer
(solid line).
Near the \Mn\ moments [$(z+z_0) \ll a$] the magnetic field is
effectively that of the nearest moment and therefore $H_z^d(z)$
follows the asymptotic behavior for a dipolar field of a single
moment, which agrees with Eq.~(\ref{Bmax}) with $\alpha=3.0$ (dotted
line). However, for increased stopping depths [$(z+z_0) \sim a$], the
field experienced by each \Li\ contains significant contributions from
several \Mn\ moments on the surface. This results in cancellations
that lead to a faster decrease in the fields (reflected by a deviation
from the $\alpha=3.0$ power law behavior). Finally, at even greater
distances the magnetic field becomes almost uniform, similar to the
case of a dipolar field from a uniform magnetic layer\cite{Xu07JMR}
(The small but non-vanishing value of $\Delta$ in Fig.~\ref{Square} is
due to the finite size of lattice used in the simulations as well as
rounding errors). In our experiment, we expect considerable randomness
in the \Mn\ arrangement in the monolayer
\cite{Condorelli06CEJ}. However, simulations show that our conclusions
from the model described above are independent of the detailed
arrangement of the \Mn\ moments, since randomness introduces only a
reduction in the asymptotic value of $\alpha$ for $(z+z_0) \ll a$ and
a slight increase of $\alpha$ and the range of the dipolar fields
deeper into the substrate, as can be seen in Fig.~\ref{Square} (dashed
line). We would like to point out here that since the implanted \Li\
senses mainly the few nearest grafted neighbours ($< 10$ molecules),
the simulations which assume a perfect flat substrate are still valid
if the surface roughness is small ($\ll 1$ nm) within the area
occupied by these neighbours. This is the case for our Si substrates,
where the roughness is $0.1-0.2$ nm over an area of at least $200
\times 200$ nm\cite{Condorelli04ACIE,Condorelli06L}.

\begin{figure}[h]
\centerline{\includegraphics[width=\columnwidth]{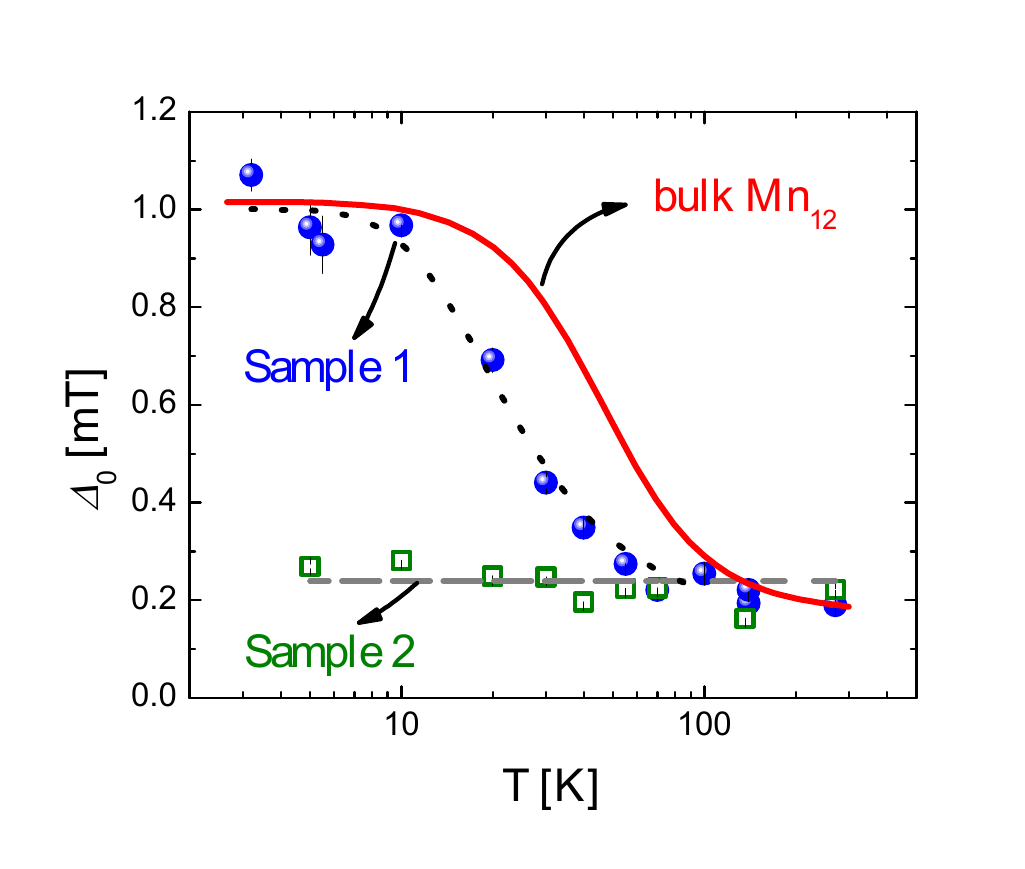}}
\caption{The measured broadening $\Delta_0$ in samples {\bf 1}
  (circles) and {\bf 2} (squares) as a function of temperature at
  $E=1$~keV. The solid line is the measured $m_z$ in bulk and the
  dotted line is a guide to the eye. The dashed line represents the
  average $\Delta_0$ measured in the control sample {\bf 2}.}
\label{BmaxvsT}
\end{figure}
In addition to the obvious broadening at low $E$, the observed
line-shape also changes. At high $E$ [bottom of Fig.~\ref{Mn12-3D}(c)]
the line-shape fits well to a simple Lorentzian function, while at low
$E$ [top of Fig.~\ref{Mn12-3D}(c)] it has a different shape
characterized by a sharp center and broad tails. The intrinsic
resonance line-shape of \Li\ in Si is that obtained at high
implantation energy. Therefore, the low implantation energy line-shape
may be simulated by calculating the broadening of the intrinsic line
due to dipolar fields generated by the monolayer. For simplicity we
assume that the magnetic field distribution at a depth $z$ in the Si
substrate, $n(B,z)$, is a uniform distribution between $\pm \Delta(z)$
[Eq.~(\ref{Bmax})]. This assumption is necessary due to the lack of
knowledge of the exact structure of the grafted \Mn\ moments lattice
on the surface, and therefore it is impossible to simulate the exact
form of $n(B,z)$. However, this allows (at least qualitatively) an
estimate of the size of dipolar fields as a function of depth. For
each implantation energy we calculate a depth-averaged field
distribution:
\begin{equation}
\langle n(B) \rangle = \int \rho(z) n(B,z) dz,
\end{equation}
where $\rho(z)$ is the stopping distribution obtained using the
TRIM.SP code \cite{TRIM,Morenzoni02NIMB} to simulate the implantation
profile of \Li\ in Si. The final step in generating the line-shape is
to convolute $\langle n(B) \rangle$ by the intrinsic Lorentzian
line-shape, i.e. the line-shape obtained from the high $E$
measurement. Recall, this line-shape is identical to that obtained in
sample {\bf 2}, but it represents a more accurate in-situ reference to
the low $E$ resonance since it can be measured at exactly the same
experimental conditions (temperature, $H_1$, etc.). The calculated
line-shape is used to fit the \bnmr\ spectra, e.g. the solid line in
Fig.~\ref{Mn12-3D}(c), where $\Delta_0$, $\alpha$ and $z_0$ are the
fitting parameters.

The best fit of the resonance lines at the implantation energy of
$E=1$~keV and all temperatures is achieved with a common $\alpha= 3.0
\pm 0.1$ and $z_0 = 1.2 \pm 0.1$~nm, while $\Delta_0$ varies with
temperature. In Fig.~\ref{BmaxvsT} we plot the fitted values of
$\Delta_0$ as a function of temperature for both samples {\bf 1}
(circles) and {\bf 2} (squares). At high temperatures the width
$\Delta_0$ is small, $\sim 0.2$~mT, and is {\em equal in both
samples}. However, in sample {\bf 1} it increases dramatically as the
temperature is lowered below $\sim 100$~K reaching $\sim 1.1$~mT at
$T=3.2$~K, while it remains unchanged in sample {\bf 2}. Clearly, this
temperature dependent broadening is due to the \Mn\ magnetic moments
at the surface of sample {\bf 1}. The small $\Delta_0$ at high
temperature in both samples is unrelated to the \Mn\ magnetic moments,
but rather it is likely due to changes in the Si structure near the
surface, caused by the grafted ligands and resulting in a small
quadrupolar broadening \cite{Salman04PRB,Salman06PB,Salman06PRL}.

As discussed above, a measurement of $\Delta_0$, or more precisely the
difference between the broadening in samples {\bf 1} and {\bf 2}, as a
function of temperature is equivalent to measuring the $z$ component
of the effective magnetic moment ($m_z$) of a single \Mn\ molecule. As
shown in Fig.~\ref{BmaxvsT}, there is a sharp increase below $\sim
100$~K and saturation at low temperature. The increase of $m_z$ below
$100$ K is indicative of the gradual de-population of thermally
activated states. The low temperature saturation occurs when most of
the \Mn\ moments reside in their ground spin state in this 2D
configuration and are aligned with the applied magnetic field. We
compare the measured magnetization for the monolayer to that measured
in a bulk \Mn\ sample at the same applied field (solid line in
Fig.~\ref{BmaxvsT}). The bulk magnetization was scaled to match the
low temperature broadening. Clearly, there is a dramatic difference
between our experimental results in the monolayer compared to that in
the bulk. This difference is a strong indication that the magnetic
properties of \Mn\ in the 2D configuration are significantly different
from the bulk. Earlier studies suggest that the \Mn\ clusters in the
monolayer remain intact \cite{Condorelli06CEJ}. Hence the difference
is most probably due to changes in their electronic structure, which
may be caused by distortions of the \Mn\ core in the monolayer due to
the different local environment.

\begin{figure}[h]
\centerline{\includegraphics[width=\columnwidth]{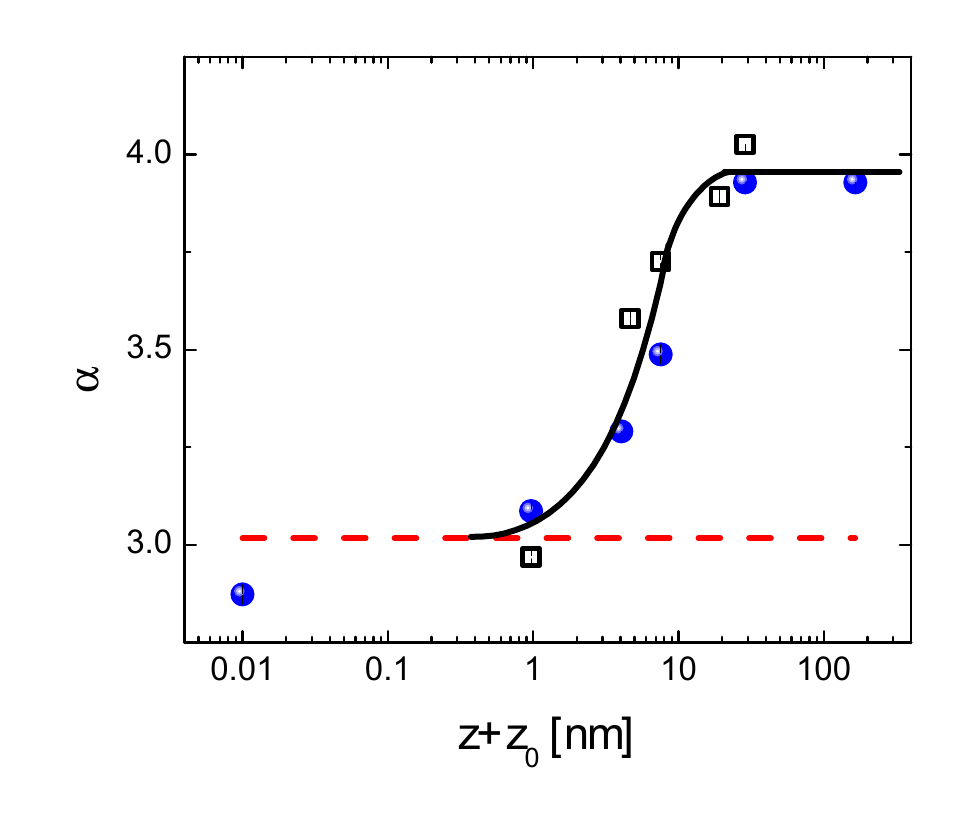}}
\caption{$\alpha$ as a function of implantation depth estimated from
  fitting the resonances at $T=3.2$~K (circles) and $5$~K (squares).
  The solid line is a guide to the eye and the dashed line represents
  the asymptotic value $\alpha=3.0$ near the monolayer.}
\label{Alphavsz}
\end{figure}

Several other points are noteworthy. No shift in the resonance
frequency is observed. Simulations show that this is expected due to
the randomness in the lattice which acts to reduce the shift to below
experimental resolution ($< 0.05$~mT). In addition the value of $z_0
=1.2 \pm 0.1$~nm, obtained from the fits, is in very good agreement
with the thickness of the grafted layer ($\sim 1.1$~nm) measured using
atomic force lithography \cite{Condorelli04ACIE}. Inspecting the
resonance lines at different depths and $T=3.2$ and $5$~K shows that
$\alpha$ (obtained from best fits using a common value of $\Delta_0$
and $z_0$ for each temperature) exhibits a strong dependence on the
implantation depth (Fig.~\ref{Alphavsz}).
As expected, we find that at low implantation depth (near the
monolayer) $\alpha \approx 3$, with a large deviation at larger
depths. The deviation from the asymptotic value $\alpha=3.0$ begins to
occur when the average implantation depth exceed $\sim 1$~nm. Compared
with simulations of the dipolar fields of the \Mn\ moments on the Si
substrate, this corresponds to $0.3a-0.5a$ and allows a rough estimate
of the average distance between neighbouring \Mn\ molecules of $a \sim
2-3.3$~nm, a reasonable value considering the size of \Mn\ molecules
core \cite{Condorelli04ACIE}. Finally, using the extracted values of
$a$ and $z_0$ to simulate the dipolar field to roughly estimate
$\Delta_0$, we find that the low temperature average of $\Delta_0 \sim
1$~mT corresponds to a \Mn\ magnetic moment of $5 \mu_B-12 \mu_B$, as
expected for an electronic magnetic moment with a large effective
spin.

In conclusion, we used \bnmr\ of \Li\ to measure the effective
magnetic moment of a single \Mn\ molecule in a monolayer grafted on
Si, demonstrating that the technique has the required sensitivity to
investigate the magnetic properties of a sub-monolayer of magnetic
molecules. The temperature dependence of the \Mn\ magnetic moment
indicates that their magnetic properties and spin Hamiltonian are
dramatically different from bulk. Since future practical applications
of SMMs will undoubtedly require them to be fabricated in the form of
monolayers, it is important to understand and thus control any
modifications that result from depositing them on
surfaces\cite{Jo06NL}.

This work was supported by the CIAR, NSERC and TRIUMF. We thank Syd
Kreitzman, Rahim Abasalti, Bassam Hitti and Donald Arseneau for
technical support.

%\bibliographystyle{apsrev}
%\bibliography{/home/zaher/LaTex/referances}
\newcommand{\noopsort}[1]{} \newcommand{\printfirst}[2]{#1}
  \newcommand{\singleletter}[1]{#1} \newcommand{\switchargs}[2]{#2#1}

\end{document}